\documentstyle[prl,aps,epsf]{revtex}
\draft
\begin{document}
\twocolumn[\hsize\textwidth\columnwidth\hsize\csname @twocolumnfalse\endcsname

\title{Pinning and Dynamics of Colloids on 
One Dimensional Periodic Potentials}   
\author{C. Reichhardt and C.J. Olson Reichhardt} 
\address{
Center for Nonlinear Studies and Theoretical Division, 
Los Alamos National Laboratory, Los Alamos, New Mexico 87545}

\date{\today}
\maketitle
\begin{abstract}

Using numerical simulations we study the pinning and
dynamics of interacting colloids on periodic one-dimensional 
substrates. As a function of colloid density,
temperature, and substrate strength, we find a variety of 
pinned and dynamic states including  
pinned smectic, pinned buckled, two-phase flow,
and moving partially ordered structures. 
We show that for increasing colloid density,
peaks in the depinning threshold occur at 
commensurate states.
The scaling of the pinning threshold versus substrate strength
changes when the colloids undergo a transition from one-dimensional
chains to a buckled configuration.
\end{abstract}

\vskip2pc]
\narrowtext

Assemblies of interacting colloidal particles  
in two dimensions (2D)  
have attracted considerable attention
as an ideal model system 
in which various types of equilibrium phases
and melting transitions
can be studied conveniently \cite{Murray}. 
Some attractive features of this system 
include the fact that the 
colloid-colloid interactions and density can be changed easily and 
that the individual colloid positions and motions are directly accessible.   
Further, when
a 1D or 2D periodic substrate is added to the system,
unique ordering and melting transitions appear. 
In experiments on colloids interacting with 1D substrates
created using interfering laser beams,
a novel laser induced freezing was observed
in which the colloids freeze into a crystal as  the
substrate strength is increased \cite{Leiderer}. This 
substrate-induced freezing effect has also been studied 
theoretically \cite{Sood} and numerically \cite{Chakrabarti}.
At even higher substrate strengths, a reentrant laser induced {\it melting} 
can occur when the laser or substrate 
strength is large enough that the colloids behave 
one-dimensionally and the fluctuations 
are effectively enhanced, 
as predicted initially in theoretical studies \cite{Chakrabarti}.  
In this case, the colloids form a smectic state.
Laser induced freezing
and melting have subsequently been studied both experimentally 
\cite{Bechinger} and theoretically \cite{Strepp,Frey}. 
A rich variety of other equilibrium phases for varied 
colloid density on 1D substrates have been 
predicted \cite{Frey} and in some cases 
experimentally confirmed \cite{Buamgartl}. 
More recently the crystalline phases and melting of colloids interacting
with 2D periodic substrates have been studied 
and a variety of novel crystalline orderings and melting phenomena 
were shown to occur
\cite{Reichhardt,Brunner,Mangold,Trizac}.     

Far less is known about the {\it dynamical} interactions of
colloids with periodic substrates. Recently it was shown that 
dynamical locking effects can occur for colloids driven over 2D
periodic substrates when the colloids preferentially move along the symmetry 
directions of the substrate \cite{Grier,MacDonald,Gopinathan}. 
These locking effects may
be potentially useful as a technique for separating particle mixtures. 
Colloids driven over periodic substrates are also a useful
model system for studying depinning phenomena. Similar systems
in which depinning on periodic substrates occurs include
vortices in superconductors interacting with 1D
and 2D periodic pinning arrays \cite{Baert} as well as
models for atomic friction \cite{Persson}.  
 
In this work we consider colloids driven over periodic
1D substrates, which to our knowledge has not been
studied previously.
We consider parameters relevant to recent
experiments on colloids interacting with periodic potentials.   
For fixed substrate strength and lattice constant, 
we find that as the density of the colloids increases,
peaks in the depinning threshold occur at various commensurate fillings. 
In general we observe a pinned regime, a disordered or plastic flow 
regime, and partially ordered moving regimes as a function of driving force. 
Several types of pinned states appear, include pinned commensurate 
lattices, pinned smectics, and
buckled pinned configurations. 
For high colloidal densities where the colloids form a smectic state,
we show that as the substrate strength decreases there is a change
in the scaling of the depinning force versus substrate
strength when the colloids
undergo a transition from 1D chains to a buckled state. 
 
We simulate a two-dimensional system of $N_c$ colloids
with periodic boundary conditions
in the $x$ and $y$-directions. 
The overdamped equation of motion for colloid $i$ is
\begin{equation}
\frac{d {\bf r}_{i}}{dt} = {\bf f}_{cc}^i +
{\bf f}_{s} + {\bf f}_{d} + {\bf f}_{T}
\end{equation} 
Here the colloid-colloid interaction force is 
${\bf f}_{cc}^i = -\sum_{j \neq i}^{N_{c}}\nabla_i V(r_{ij})$, where 
we use a Yukawa or screened Coulomb
potential given by
$V(r_{ij}) = (q_{i}q_{j}/|{\bf r}_{i} - {\bf r}_{j}|)
\exp(-\kappa|{\bf r}_{i} - {\bf r}_{j}|)$.
$q_{i(j)}=1$ is the colloid charge, $1/\kappa$ is the 
screening length, and ${\bf r}_{i(j)}$ is the position of 
particle $i (j)$. 
The substrate is periodic in 1D as shown in Fig. 1(a),
with period $d$ in the $y$-direction,  
${\bf f}_{s} = f_p\sin{(2\pi y/d)}{\bf {\hat y}}$. This
is the form expected from the modulated laser fields 
used in the experiments \cite{Bechinger}.    

\begin{figure}
\center{
\epsfxsize=3.5in
\epsfbox{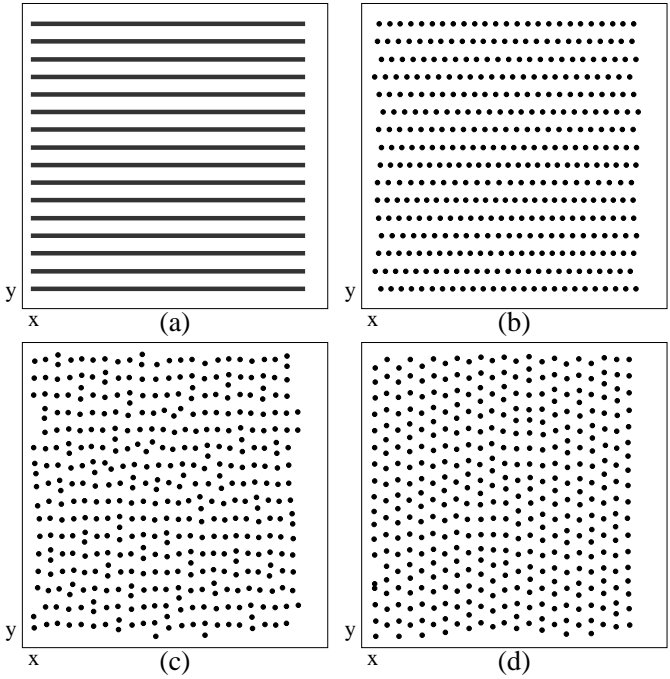}}
\caption{
(a) Heavy lines indicate the locations of the substrate minima.
(b-d) Colloid configurations (black dots) for 
a system with fixed $d$, fixed $f_{p}=2.0$, and $n_{c} = 1.35$. 
(b) $f_{d} = 0.0$ where a pinned smectic state occurs.
(c) $f_{d}/f_{c} = 1.1$, where $f_{c}$ is
the depinning threshold. The colloids are moving in the $y$ direction.
(d) $f_{d}/f_{c} = 1.2$. Here the colloids show
significant triangular ordering. 
}
\end{figure}

\noindent
The external driving force is  ${\bf f}_{d}=f_d {\bf {\hat y}}$ which could
come from an applied electric field. 
The thermal force ${\bf f}_{T}$ is a randomly fluctuating force 
from Langevin kicks. 
We measure lengths in units of $d$, the substrate periodicity.
We have considered various system sizes; here we focus on 
fixed $L_{y} =  16d$. A triangular colloidal lattice 
is commensurate with the substrate
for $d = \sqrt{3}a/2$, where $a$ is the colloidal
lattice constant. The density of colloids, $n_{c}$, is normalized
such that at the commensurate density, $n_{c} = 1.0$. 
Temperature is reported in terms of the 
melting transition temperature
$T_m$ for a colloidal lattice at the commensurate 
density in the absence of the substrate. 
For the results in this work we focus on the case $T/T_{m} = 0.5$
which is low enough to avoid appreciable creep.   
The initial colloidal state is obtained at zero external drive
by cooling the system from a high temperature molten state
to a lower temperature. 
A similar procedure was used to find the equilibrium states for
colloids on 2D substrates \cite{Reichhardt}.  
The applied drive is then increased
from $f_{d} = 0.0$ by small increments. At each increment, 
we measure the average colloidal velocity 
$V_y=\sum_{i=1}^{N_c}{\bf v}_i \cdot {\hat y}$ after reaching the
steady flow state.

We first consider the case of fixed substrate 
lattice constant $d$ and strength $f_p$ 
and examine the depinning as a function of colloidal
density $n_c$. In general we find three 

\begin{figure}
\center{
\epsfxsize=3.5in
\epsfbox{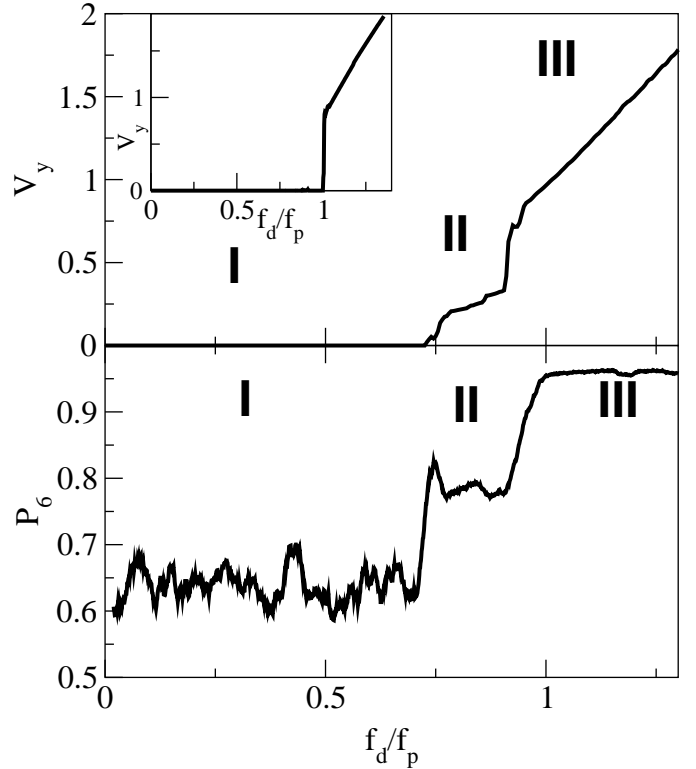}}
\caption{ (a) Average colloidal velocity $V_y$ vs $f_{d}/f_{p}$ for the system
in Fig.~1 with $f_p=2$ and $n_c=1.35$. 
Inset: $V_y$ vs $f_{d}/f_{p}$ for $n_{c} =  0.375$.
(b) The density of six-fold coordinated colloids $P_{6}$  vs $f_{d}/f_{p}$
for the system in Fig.~1.       
}
\end{figure}

\noindent
regimes: pinned,
disordered flow, and a partially ordered high drive flow. 
The exact type of order in the pinned regime depends on the 
colloid density. For high densities $n_{c} > 1.0$, the colloids
form a pinned smectic state, as seen in Fig.~1(b) 
for $n_{c} = 1.35$. 
At most densities, including this one,
the colloid lattice is not commensurate with the substrate, 
so the potential minima capture differing numbers of particles and
there are dislocations present in the lattice
with Burgers vectors
aligned parallel to the substrate.
These dislocations 
depin before the particles do
as the drive is increased so the colloids generally 
initially depin into a plastic flow regime or two phase flow 
regime where only a 
portion of the colloids move. 
In Fig.~1(c) we show the colloidal
positions just above depinning for $f_{d}/f_{c} = 1.1$, where 
$f_{c}$ is the critical depinning force, 
for the same system in Fig.~1(b). Here the colloidal positions are
much more disordered, and a portion of the
colloids are pinned while the other portion is moving.
As $f_{d}$ is further increased, all the colloids
depin and form a moving partially
ordered state where the colloids regain a considerable amount of 
triangular ordering, as seen in Fig.~1(d).  The colloid lattice
also realigns in the direction of the drive.

The different flow regimes can be identified by 
the amount of disorder in the
lattice and the characteristics of the colloid velocity vs external
force curves. 
In 

\begin{figure} 
\center{
\epsfxsize=3.5in
\epsfbox{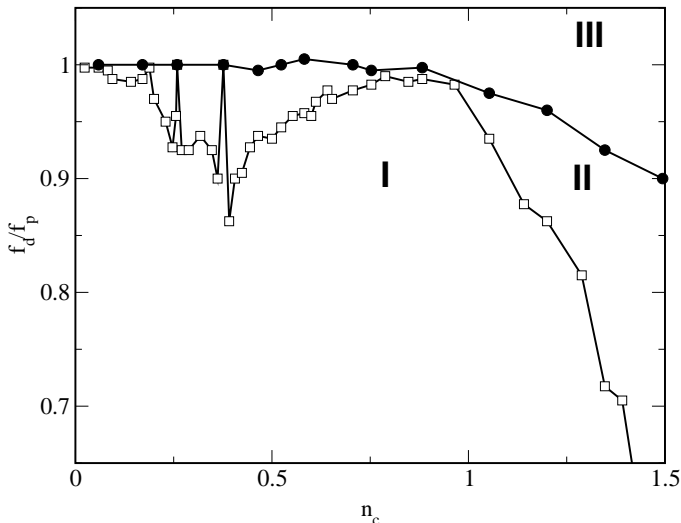}}
\caption{
Flow regions I (pinned), II (plastic), and III (partially ordered)
for $f_{d}/f_{p}$ vs colloid density 
$n_{c}$ at constant $f_p=2.0$. Open squares: 
depinning threshold. Filled circles: reordering crossover.  
}
\end{figure}

\noindent
Fig.~2(a) we plot the average colloid velocity $V_y$ 
and 
in Fig.~2(b) we show the corresponding fraction of $6$-fold 
coordinated particles $P_{6}$ 
vs the external drive $f_{d}/f_{p}$
for the system with $n_{c} = 1.35$. 
For a triangular lattice, $P_{6} = 1.0$.   
In Fig.~2(a), there are three regions of the velocity force curve
labeled I, II and III.  Region I is 
a pinned state for $f_{d}/f_{p} < 0.75$.
The plastic flow region II occurs  
for $0.75 <f_{d}/f_{p} < 0.95$ and
begins when $V_y$ jumps above zero.
In the plastic flow regime, $V_y$ is smaller than 
it would be if all the colloids were moving, and lies below
the value that would be obtained by a linear extrapolation
of the velocity-force slope at higher drives. 
For $f_{d}/f_{p} > 0.94$, there is a final jump in $V_y$ indicating
that more colloids are now moving, and
the system enters region III where $V_y$ increases linearly with $f_{d}$ and 
all the colloids move with a uniform velocity. 
The three regions can also be distinguished in $P_{6}$. In the pinned 
region I where colloids are smectically ordered, there are
a significant number of dislocations giving $P_{6} = 0.65$.
In region II flow, 
a number of the pinned dislocations 
annihilate; thus, the average coordination number increases to
$P_{6}  = 0.78$. In region III, the 
colloids form a mostly triangular lattice, giving 
$P_{6} = 0.95$. In the inset of Fig.~2(a) 
we show the velocity force characteristics  
for a system with $n_{c} = 0.375$ 
where the lattice makes a transition directly from a pinned
triangular lattice to a moving triangular lattice. Here the
intermediate jumplike features in $V_y$ are missing due to the absence of 
a plastic flow regime. 

In Fig.~3 we map regions I through III 
as a function of $f_{d}/f_{p}$ vs $n_{c}$ 
for fixed $f_{p}=2.0$ and fixed $d$.
The depinning threshold shows several peaks and a prominent
broad maximum between $0.75 < n_{c} < 1.0$, corresponding to 
a triangular colloidal lattice.
The peaks occur at 

\begin{figure}
\center{
\epsfxsize=3.5in
\epsfbox{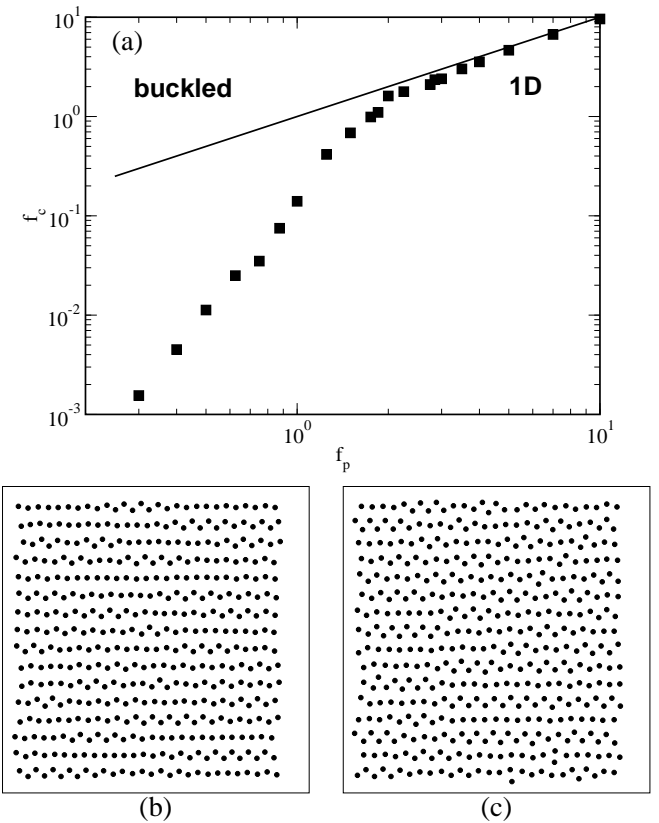}}
\caption{(a) Squares: critical depinning force $f_{c}$ vs $f_{p}$ for fixed
$n_{c} = 1.35$.  Solid line: $f_{c} = f_{p}$ curve.    
(b) Buckled state at $f_p=1.0$ and zero drive.  (c) Buckled state at
$f_p=0.75$ and zero drive.
}
\end{figure}

\noindent
commensurate densities 
$n_{c} = 0.38, 0.26$ and $0.185$, corresponding to 
a colloidal lattice constant of
$a = \sqrt 2 d$, $\sqrt 3 d$,and $2d$, respectively.  
At these densities there are no dislocations present in the system 
and the colloids pass directly from a pinned
triangular solid to a moving triangular lattice 
as a function of driving force without an intermediate
disordered flow regime.  
Region II persists in a small sliver between regions I and III near
the commensurability at $n_c=1$ due to the fact that the 
system was not perfectly commensurate and there were a
small number of dislocations present.
There is a broad minimum in the depinning threshold centered around
$n_c=0.45$.  Over this range of density
there are significant numbers of dislocations  present in the system.
For $n_{c} > 1.0$ the depinning threshold drops 
dramatically with density as dislocations proliferate
and the system enters a pinned smectic state. 
For all densities, 
once the drive is large enough for all the colloids to depin,
the system enters the partially ordered flow regime III. 
The region III boundary roughly coincides with $f_d=f_{p}$, 
although for $n_{c} > 1.0$, the onset of
region III shifts slightly down in drive for increasing $n_{c}$ due to the
enhanced colloidal interactions at these higher densities. 

The appearance of the three dynamic regimes is similar to what 
is seen for vortices driven over 2D random disorder \cite{Giamarchi}.
In the vortex case,
the plastic flow regime occurs when some vortices are trapped in
pinning sites while
additional interstitial vortices move between the pins. 
For strong driving, the vortices
form a moving smectic state aligned with the drive,
due to the effective transverse pinning which 
is present even at high drives.     
In the 1D periodic pinning case considered here, 
the partially ordered regime at high drives is a 
polycrystalline moving solid  rather than a smectic solid
since there is no transverse pinning.  

In experiments with 1D periodic arrays, it is very straightforward to
control the substrate strength by varying the laser intensity.  Thus
we consider the effects of altering the substrate force $f_p$, and
find that when the substrate is weakened for $n_{c} > 1.0$, 
a crossover occurs from the 1D pinned smectic state illustrated in Fig.~1(a)
to a buckled configuration. In Fig.~4(a) we plot the critical depinning force
$f_{c}$ vs $f_{p}$ for a system with fixed $n_{c} = 1.35$. 
The straight line in Fig.~4(a) indicates $f_{c} = f_{p}$.
For $f_{p} >1.9$ the pinned state is a one-dimensional smectic structure. 
As $f_{p}$ is further increased, $f_{c}$ increases linearly with 
$f_{p}$ and the same pinned smectic state forms. 
For $f_{p} \leq 1.9$, patches of the pinned smectic 
undergo a buckling transition where
the colloids along a single row splay out in a staggered manner,
as illustrated in Fig.~4(b) for $f_{p} = 1.0$. Here portions of the lattice
remain in the 1D state and coexist with the buckled state. 
In general the buckled portions do not form adjacent to one another
due to the increased inter-row colloidal repulsion that occurs when
the buckled state forms. 
The buckling appears at locations of enhanced stress, where dislocations in the
pattern would be present in the 1D pinned case. 
As $f_{p}$ is further decreased the buckled areas grow, as
seen in Fig.~4(c) for $f_{p} =  0.75$. The onset of the buckled state
coincides with a change from the linear behavior of $f_{c}$ with $f_{p}$
to the much more rapid nonlinear decrease for $f_{p} < 1.9$. The buckled
structure is much more weakly pinned than the 1D state since 
the force acting to push a colloid over the substrate barrier has
an additional contribution from the staggered colloids on either
side.
For pinning forces at which a buckled pinned state
forms, there is both a plastic flow regime and a crossover to 
region III flow as $f_{d}$ approaches $f_{p}$.  

In conclusion, we have numerically studied the dynamics 
and pinning 
of colloids driven over one-dimensional periodic substrates.
We find three general regimes, pinned, disordered, and partially ordered,
which can be characterized by the amount of disorder
in the colloidal lattice and by 
features in the velocity force curves. We map 
these regimes as a function of colloid density and show that peaks in the 
depinning threshold occur at commensurate densities where 
triangular or mostly triangular colloid lattices form.
As the colloidal density 
increases for strong substrate strengths,
the colloids form a pinned smectic state  
which transforms  to a buckled structure as the substrate
strength is reduced. In the pinned smectic state, the depinning threshold
decreases linearly with decreasing substrate strength, while in the 
buckled state, the depinning threshold decreases much faster than linearly
with substrate strength. 

We thank C.~Bechinger and M.B.~Hastings 
for useful discussions. 
This work was supported by the U.S. DoE under Contract No.
W-7405-ENG-36.



\begin{references}

\bibitem{Murray}

C.A.~Murray, W.O.~Sprenger, and R.A.~Wenk, Phys.~Rev.~B {\bf 42}, 688 (1990); 
K.~Zahn, R.~Lenke, and G. Maret, Phys.~Rev.~Lett.~{\bf 82}, 2721  (1999). 

\bibitem{Leiderer}
A.~Chowdhury, B.J.~Ackerson, and N.A.~Clark, Phys.~Rev.~Lett.~{\bf 55}, 833 
(1985).   

\bibitem{Sood}
J.~Chakrabarti, H.R.~Krishnamurthy, and A.K.~Sood, Phys.~Rev.~Lett.~{\bf 73},
2923 (1994). 

\bibitem{Chakrabarti}
J.~Chakrabarti, H.R.~Krishnamurthy, A.K.~Sood, and S.~Sengupta, 
Phys.~Rev.~Lett.~{\bf 75}, 2232 (1995). 

\bibitem{Bechinger}
Q.-H. Wei, C.~Bechinger, D.~Rudhardt, and 
P.~Leiderer, Phys.~Rev.~Lett.~{\bf 81}, 2606 (1998);
C.~Bechinger, M.~Brunner, and P.~Leiderer, Phys.~Rev.~Lett.~{\bf 86},
930 (2001). 

\bibitem{Strepp}
W.~Strepp, S.~Sengupta, and P.~Nielaba,
Phys.~Rev.~E {\bf 66}, 056109 (2002). 

\bibitem{Nelson} 
E.~Frey, D.R.~Nelson, and L.~Radzihovsky, Phys.~Rev.~Lett.~{\bf 83},
2977 (1999). 

\bibitem{Frey}
L.~Radzihovsky, E.~Frey, and D.R.~Nelson, Phys.~Rev.~E {\bf 63}, 
031503 (2001). 

\bibitem{Buamgartl}
J.~Baumgartl, M.~Brunner, and C.~Bechinger, Phys.~Rev.~Lett., in press. 

\bibitem{Reichhardt}
C. Reichhardt and C.J.~Olson, Phys.~Rev.~Lett.~{\bf 88}, 248301 (2002). 

\bibitem{Brunner}
M.~Brunner and C.~Bechinger, Phys.~Rev.~Lett.~{\bf 88}, 248302 (2002). 

\bibitem{Mangold}
K.~Mangold, P.~Leiderer, and C.~Bechinger,
Phys.~Rev. Lett. {\bf 90}, 158302 (2003). 

\bibitem{Trizac}
R.~Agra, F.~van Wijland, and E.~Trizac, Phys.~Rev.~Lett.~{\bf 93}, 
18304 (2004). 

\bibitem{Grier}
P.T.~Korda, M.B.~Taylor, and D.G.~Grier,
Phys.~Rev.~Lett.~{\bf 89}, 128301 (2002). 

\bibitem{MacDonald} M.P.~MacDonald, G.C.~Spalding, and K.~Dholakia,
Nature {\bf 426}, 421 (2003). 

\bibitem{Gopinathan}
A.~Gopinathan and D.G.~Grier, Phys.~Rev.~Lett.~{\bf 92}, 130602 (2004). 

\bibitem{Baert}
M.~Baert, V.V. Metlushko, R. Jonckheere, V.V. Moshchalkov, and
Y. Bruynseraede,
Phys.~Rev.~Lett.~{\bf 74}, 3269 (1995);
A.N.~Grigorenko {\it et al.}, Phys.~Rev.~Lett.~{\bf 90}, 237001 (2003).

\bibitem{Persson}
B.N.J.~Persson, {\it Sliding Friction: Physical Principles and Applications}
(Springer, Heidelberg, 2000), 2nd ed.

\bibitem{Giamarchi}
L.~Balents, M.C.~Marchetti and L.~Radzihovsky, 
Phys.~Rev.~Lett.~{\bf 78}, 751 (1997); Phys.~Rev B {\bf 57}, 7705 (1998); 
P. Le Doussal and T. Giamarchi, Phys.~Rev.~B {\bf 57}, 11356 (1998). 

\end{references}
\end{document}